# Analytical Study of Thermal Annealing Behaviour of Erbium Emission in $Er_2O_3$-Sol-Gel Silica Films

S Abedrabbo[1,2,3], B Lahlouh[1] and A T Fiory[2]

[1] Department of Physics, University of Jordan, Amman, Jordan 11942
[2] Department of Physics, New Jersey Institute of Technology, Newark, NJ 07901

**Abstract**

Room-temperature 1535-nm-band photoluminescence in ~126 nm silica films (6 at. % doping), produced by spin-coating an $Er_2O_3$ and tetraethylorthosilicate sol-gel formulation on silicon substrates, was studied as a function of vacuum furnace annealing (500 to 1050 °C). Emission is strongly enhanced for annealing near 850 °C, which is shown by modeling the temperature dependence as arising from thermally-activated removal of hydroxyl ions. Suitability of such a process for silicon-based applications is discussed.

## 1. Introduction

Rare-earth doped optical materials are commonly used in optical telecommunication networks for applications such as optical amplifiers, waveguides, infrared light sources, integrated optical devices, and lasers [1-5]. Erbium (Er) doped silica ($SiO_2$:Er), which is the most widely used optical material for these purposes, can be prepared in film form by a variety of methods, although sol-gel spin coating has received special attention, owing to low deposition temperature and low overall cost [3,4,6-10]. The sol-gel technique developed for this work uses erbium oxide ($Er_2O_3$) and high Er concentration (6 at.%) in the $SiO_2$:Er film, which yields optically active films at moderate annealing temperatures (~850 °C); suitable applications are e.g. planar erbium-doped waveguide amplifiers (EDWAs).

Erbium-doped silica is obtained by hydrolysis of $Si(OC_2H_5)_4$ (tetraethylorthosilicate or TEOS) in a solution containing an erbium compound, such as its oxide in the present study, or as a salt and then annealing the gel. Special considerations of sol-gel processes concern quenching of the 1535-nm emission (Stark-split intra-4f $^4I_{13/2}$ - $^4I_{15/2}$ transitions [1]) by resonant energy transfer to vibrations of residual OH [11,12] and dependence of external emission on Er concentration [7,8]. Co-doping the silica with trivalent ions such as $Al^{3+}$ and $Ti^{3+}$ has been used to promote solubility and suppress clustering of the rare earths [3,4,7,13-15], or by using additives such as $P_2O_5$ that act as glass stabilizers [6,16]. Utilizing these additives suggests an upper limit of about 1 at.% in Er concentration, owing to observed concentration quenching upon thermal annealing [7]. The process described in this work, using the relatively high Er concentration of 6 at.%, is expected to yield Er-O:Si-O structures that inhibit precipitation, owing to the strong Er-O bond (6.3 eV [17]) and the role of Er as a network element in glasses [18,19].

---

[3] E-mail: sxa0215@yahoo.com

Although hydroxyl contamination appears to be virtually eliminated for high temperature annealing (900 – 1100°C) [5,20], process temperatures above 1100 °C can lead to devitrification on a nanocrystalline scale [21]. Moreover, facile integration with silicon-based optoelectronics prefers processes with annealing below 1000 °C. This tends to narrow applicability of polycrystalline materials, such as the Er silicates that optimize at high temperatures (> 1100 °C) [22,23]. This presents the challenge, which is studied in this work, of adequately removing the effects of residual hydroxyls in sol-gel films [5] without the need for high temperature processing.

Section 2 presents the method of processing the sol-gel films for this study, including their characterization by optical measurements of thickness, spectral indices of refraction and photoluminescence. Results are presented in section 3; a reaction-rate analysis of the dependence of photoluminescence on annealing temperature is presented in section 4; interpretation of the results in terms of OH removal and concentration quenching are discussed in section 5; and conclusions are presented in section 6.

## 2. Experimental procedure

The sol-gel process in this work includes the following distinctive features: (1) Er is introduced into the sol as $Er_2O_3$ powder with acetic acid catalysis, thereby increasing the available oxygen while also limiting the presence of foreign atoms in the $SiO_2$:Er film to only hydrogen and carbon; (2) Er is incorporated at high concentration yielding about 6 at. % Er in the $SiO_2$:Er film; and (3) other glass-promoting and stabilizing additives are not used.

The starting precursor solution was prepared by mixing 0.5 g $Er_2O_3$ powder into a solution of 4ml ethanol, 4 ml acetic acid, and 1.6 ml deionized water that was stirred at 45 °C for 3 hours. The sol solution was then prepared by the addition of 2 ml TEOS and stirring for 10 min at 80 °C. Hydrolysis with high water/TEOS molar ratio R (here, R = 10) with acid catalysis produces branched Si-O polymerization in the sol and leads to dense films [24]; the trade-off is increased hydroxyl content to be removed by annealing [10]. Following this step, the solution was passed through a syringe filter with 0.45-µm pore size. The filtered solution was then spun coated on 2.5-cm pieces of cleaned Si (100) silicon substrates rotating at 1200 rpm for 30 seconds. The resulting gel films were then oven dried in air at 120 °C for 30 minutes.

Post-deposition thermal treatments studied in this work comprise vacuum annealing at temperatures from 500 to 1050 °C, which are comparable to those in previous studies of Er-salt sol-gel processing [7,9,10,16]. A set of ten samples were annealed for one hour under moderate vacuum of 2 Pa at temperatures in the range of 500 to 950 °C in a single-stage horizontal vacuum furnace. A second set of similarly prepared samples were annealed at 1050 °C in air and in vacuum. Prior work has shown that vacuum annealing is more effective in removing OH contaminants than annealing in air, as determined from increased $Er^{+3}$ emission lifetime [10]. The choice of annealing temperatures also takes into consideration thermal reflow processes for deposited silica dielectrics used in silicon microelectronics fabrication, which is preferably 1000 °C or lower [25].

Film thickness and index of refraction as a function of wavelength in the region 240 - 860 nm were determined using a spectrophotometer designed for thin-film metrology (FilmTek 3000 [26]). A model Fluorolog-3 spectrofluorometer (Horiba Jobin Yvon) was used to obtain room temperature photoluminescence (PL) emission from the $Er^{3+}$ centers in the 1535 nm band. A Xe lamp was used for excitation with a double excitation monochromator to set a fixed excitation wavelength in the range of 515 to 530 nm. Spectral signal intensities were recorded in the range 1400 to 1600 nm using a cooled Hamamatsu InGaAs photodiode detector, preceded by a single emission monochromator, using 0.2 s integration time at each wavelength. Emission in the 1535 nm band was maximized by tuning the excitation monochrometer to 521 – 523 nm, as obtained from a three-dimensional matrix of photoluminescence-excitation spectra. Photoluminescence spectra were normalized to the power output of the excitation source, monitored by a separate photodiode.

## 3. Experimental results

Thermal annealing has the effect of reducing film thickness *d* and increasing the index of refraction *n*. Figure 1 shows representative data for an as-deposited (air dried gel) film and three films annealed at temperatures of 600, 650 and 700 °C (*n* is reproducible to ± 0.029 at 632 nm for air dried films). Corresponding film thicknesses of 168, 138, 138, and 126 nm, respectively, were determined simultaneously by the spectrophotometer. Spectral index of refraction $n(\lambda)$ increases and film thickness decreases with annealing temperatures above 500 °C, owing to the densification of sol-gel films, primarily by removal of OH, decreasing porosity, and enhanced Si-O (and presumably also Er-O) bonding. The wavelength dependence of $n(\lambda)$ in annealed films has a form similar (albeit with reduced $n$) to that of silica. Index $n$ approaches the value of 1.46 at the benchmark wavelength of 632 nm, as found in previous studies of annealed Er-doped TEOS-based sol-gel processes; for higher temperature annealing, $n(\lambda)$ is similar to the 700 °C data in figure 1 [3,4,10].

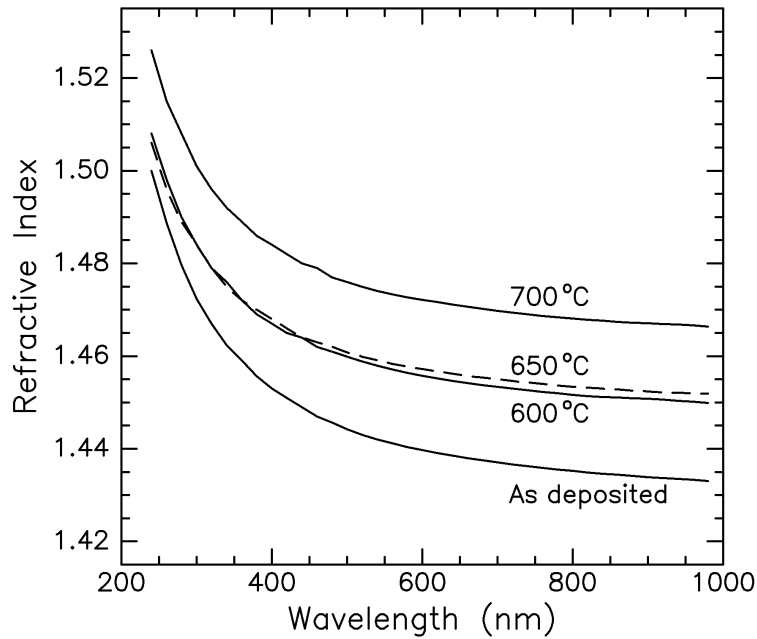

**Figure 1.** Refractive index spectra of Er-doped sol-gel silica films for as-deposited (post bake) and three annealing temperatures.

Photoluminescence spectra of the annealed samples are presented in figure 2 (normalized intensity scale). Emission from the as-deposited sample is at best feeble and barely noticed when compared to annealed samples. This is to be expected, since low temperature baking leaves residues from the sol and an abundance of water and hydroxyls in a low-density porous near-glassy network. As such it is expected that non-radiative recombinations will strongly compete with the radiative ones, leading to the low overall PL signal observed.

The PL signal appreciably improves as a function of increasing annealing temperature, particularly above 700 °C, until it reaches a maximum at 850 °C. Samples annealed at higher temperatures exhibit a decreasing trend in their PL (confirmed out to 1050 °C for the separately prepared samples annealed in either air or vacuum). Maximum PL intensity occurs at emission wavelengths in the vicinity of 1533 to 1537 nm and spectral full width at half maximum is in the range 51 to 58 nm, both varying somewhat among the samples. To illustrate the trend peak PL intensities (normalized arbitrary units; integrated spectra are similar) for the 500- to 950-°C anneals are plotted against annealing temperature in figure 3 as filled circles (uncertainties for anneals above 600 °C are approximately 12% full scale). The evident non-monotonicity in temperature dependence of the data is indicative of competing thermal reactions that activate optical emission from the intra 4f $Er^{+3}$ band as the annealing temperature is increased towards 850 °C, and de-activate it for annealing at the higher temperatures. In the next section we introduce a model that provides a quantitative analysis for understanding annealing behaviour.

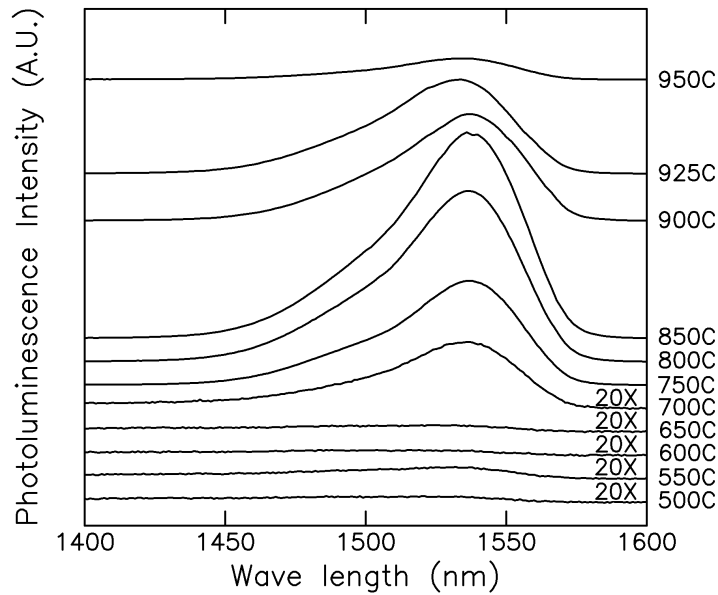

**Figure 2.** Photoluminescence spectra at the temperatures indicated; ordinates offset for clarity; lower five spectra expanded by factor 20.

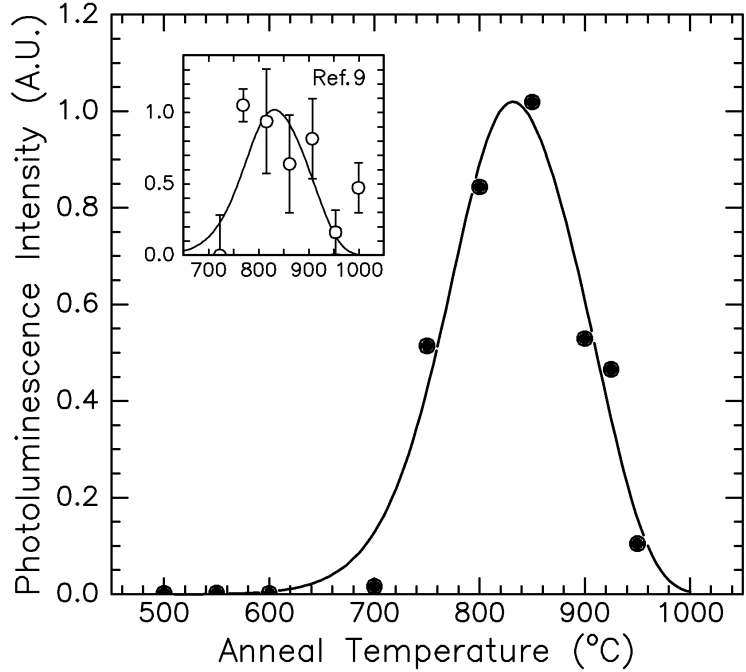

**Figure 3.** Peak photoluminescence (1535 nm band) vs annealing temperature. Points: data; curve: model function of activated fraction, equation (3), fitted to data. Inset: normalized PL *vs.* anneal temperature (circles) for Er-implanted silica colloids from [9] and same model function (curve).

**4. Thermal annealing analysis**

Study of annealing behaviour can help elucidate the underlying mechanisms for producing the optically active $Er^{+3}$ centers in $SiO_2$:Er films. There is usually an optimum annealing temperature (or range of temperatures) that depends on process method as well as Er concentration, e.g. temperatures near 900 °C are often used for annealing sol-gel films [3,4,15]. However, detailed temperature dependence is rather overlooked for Er-doped sol-gel films, in contrast to numerous studies of related film materials (e.g. Er-doped silicon suboxides [27,28] and Er-implanted silica colloids [9]). In the case of sol-gel processing, annealing is required to densify the film by removing water, organic compounds and hydroxyl residues [5,20]. Thermogravimetric analysis shows that the majority of weight loss from outgassing of volatiles occurs at temperatures below 500 °C [7]. The decrease in emission at high annealing temperatures can arise from a number causes: Segregation by precipitation of Er or phase separation by clustering, and quenching by Er-Er interactions at high optically active $Er^{+3}$ concentrations, as well as residual OH contamination [29].

To provide a tractable treatment of thermal annealing, we employ an empirical model containing two main thermal reaction processes: Formation of optically active $Er^{+3}$ is assumed to proceed with a reaction rate $r_a$; and de-activation, or quenching of optical activity, to proceed with a separate reaction rate $r_d$. A fixed fraction $f_0$ of the Er is assumed be available for optical activation, which in this model is initially optically inactive (i.e. prior to annealing). During the course of annealing a fraction $f_a$ becomes optically activated and a fraction $f_d$ becomes optically deactivated. The reactions involved are considered to transform $f_0$ into $f_a$ and $f_d$. The rate of formation of active Er is determined by the remaining fraction of available Er, which is $f_0$ less the fractions activated and deactivated, and is determined from the rate equation,

$$\frac{df_a}{dt} = r_a (f_0 - f_a - f_d). \tag{1}$$

The rate of deactivation is determined by $f_0$ less the fraction deactivated, determined as,

$$\frac{df_d}{dt} = r_d (f_0 - f_d). \tag{2}$$

The initial conditions ($t = 0$) are $f_a = f_d = 0$ in this approximation (i.e. neglecting the small initial optically active fraction). In the limit $t \to \infty$ the asymptotes are $f_a = 0$ and $f_d = f_0$, as all of the available Er is ultimately deactivated by returning to an optically inactive status (neglecting a small fraction that could remain active). The solutions to equations (1) and (2) are thus obtained as

$$f_a = \frac{r_a}{r_a - r_d} \left[ \exp(-r_d t) - \exp(-r_a t) \right] f_0, \tag{3}$$

and

$$f_d = \left[ 1 - \exp(-r_d t) \right] f_0. \tag{4}$$

Relating this to the experiment, $t$ is the annealing time ($t_{anneal} = 3600$ s) wherein temperature is held constant. Temperature dependence is introduced as Arrhenius equations for the reaction rates,

$$r_a = r_{a0} \exp(-E_a / k_B T); \tag{5a}$$

$$r_d = r_{d0} \exp(-E_d / k_B T), \tag{5b}$$

with respective prefactors $r_{a0}$ and $r_{d0}$ and activation energies $E_a$ and $E_d$. Here, $T$ is the annealing temperature.

The activated fraction $f_a$ may exhibit a maximum in its temperature dependence, which can be estimated for $r_a \gg r_d$ by the approximation,

$$T_{max} \approx \frac{E_a / k_B}{\ln[r_{a0} t / \ln(r_a / r_d)]} \quad . \tag{6}$$

This transcendental equation contains weak logarithmic dependence on $T_{max}$ in the denominator of the right hand side.

Application of this model to the data for annealing of Er-doped sol-gel films presented in section 3 involves determining the four parameters of equations (5a) and (5b) by fitting the model to the temperature dependence of the photoluminescence intensity, as given by figure 3. Of particular interest are the relative magnitudes of $E_a$, which is the activation energy of the favourable process (forming optically active $Er^{+3}$) and $E_d$, which is the activation energy of the unfavourable process (quenching optically active $Er^{+3}$). We find that $E_a < E_d$ and $r_a > r_d$ for the furnace annealing method of this work, as required for self-consistency in the model assumptions.

A non-linear least squares fit of the function $f_a$ from equation (3) to the data in figure 3, using expressions for the rates $r_a$ and $r_d$ given by equations (5a) and (5b), yields the fitted parameters $E_a = 2.3 \pm 0.6$ eV and $E_d = 2.6 \pm 0.7$ eV, with $r_{a0} \approx r_{d0} \sim 3 \times 10^7$ s$^{-1}$. The fraction $f_0$ is on the order of unity (in terms of arbitrary units of normalized PL data). The fitted function is shown as the solid curve in figure 3. While the fit confirms the premise of the model, i.e. one finds $E_a < E_d$, the activation energies themselves are subject to significant errors, owing to parameter correlation and uncertainty in the data. The magnitude of the activation energies is consistent with the idea that OH removal is the mechanism by which the otherwise optically active $Er^{+3}$ yields the PL signal. One notes that the activation energy for OH diffusion in silica glass was measured to be $E_{D,OH} = 2.46$ eV [30], which coincidentally is the same order as the fitted $E_d$. Perhaps more significantly, the fit finds that $E_d$ is comparable to $E_a$. Interpretation of these results are discussed below.

## 5. Discussion

It is known that OH is extremely effective in quenching excited $Er^{+3}$ ions [11]. At high OH concentrations direct resonant energy transfer from the excited ion to OH serves as an ~~extremely~~ effective trap. At low OH concentrations fast transfer of energy from ion to ion via the Förster mechanism (cooperative energy transfer or CET) may allow diffusion of the excitation to an $OH^-$ impurity, where it becomes dissipated by multi-phonon assisted decay to the oxide host [31]. The latter process is a type of concentration quenching, since the rare-earth concentration is the dominant factor in the CET coefficient [29]. In the present case both mechanisms are expected to apply, since the Er concentration is quite high. Hydroxyl contamination in Er-doped sol-gel films is progressively reduced with increasing anneal temperature, falling below $\sim 10^{-4}$ molar fraction, the detection limit of available IR spectrophotometers (sensitivity for absorption at 3400 – 3700 cm$^{-1}$ [5]) for ~900-°C anneals [5,20]. However, as discussed by Slooff et al. [10] $10^{-4}$ OH levels, e.g. produced by an R = 4 sol-gel process with a 900-°C anneal [4,8], are readily detected by the effects on $Er^{+3}$ lifetime [29].

Weak PL emission for the low temperature anneals ($T \ll 850$ °C) indicates quenching of excited $Er^{+3}$ ions by direct Er→OH interactions. Increase in PL intensity with annealing temperature in the region $T \leq 850$ °C is therefore consistent with removing OH from proximity to $Er^{+3}$ ions. Residual OH is mostly surface bonded within pores in the form of silanol groups, as shown by infrared absorption near 3670 cm$^{-1}$ observed in bulk silica gels (~ 900 °C anneals) [32,33]. An identified mechanism for OH removal is surface desorption of H$_2$O, which is released upon reaction of surface silanol to form siloxane bridge structures, corresponding to an enthalpy $\Delta H$ = 2.16 eV [34]. Another aspect of OH activity to consider is its diffusivity, which can be estimated from bulk studies as $D_{OH} \approx 2.7 \times 10^{-10}$ cm$^2$s$^{-1}$ at 850 °C [30]. The corresponding diffusion length $x_D = (4D_{OH} t_{anneal})^{1/2} =$ 2mm turns out to considerably exceed the film thickness (126 nm). Thus it is consistent to associate the rate $r_a$ with OH removal by desorption and to relate $E_a$ to $\Delta H$, given also that the two energies are of comparable magnitudes.

Since the observed diminution of PL intensity at higher annealing temperatures corresponds to a thermal activation energy $E_d$ that is also on the order of that for OH processes, it appears that OH removal is involved in the PL deactivation as well. Reduced concentrations of residual OH obtained for high temperature annealing may allow the diffusive Er→Er→OH processes to become effective for quenching PL emission. Moreover, high Er concentration in itself (e.g. upon virtually eliminating OH and densifying as Er-Er distances decrease) can lead to concentration quenching by cross relaxation or cooperative upconversion processes (CUP) involving Er-Er dipolar interactions, a form of CET, that typically converts one out of two units of excitation energy into heat; this CUP mechanism has been reported for optical fibers with high Er concentrations [35]. Either of these mechanisms can account for the decrease in PL emission for anneals in the region 850 °C << $T$ ≤ 1050 °C.

Diffusivity of Er at low concentrations in various deposited silica films was recently measured in the temperature range 1000 – 1100 °C [36], from which an activation energy $E_{Er}$ = 5.3 eV was derived; extrapolation to 950 °C yields diffusivity $D_{Er} \approx 6 \times 10^{-17}$ cm$^2$s$^{-1}$ and diffusion length $(4D_{Er} t_{anneal})^{1/2}$ ~ 10 nm. Erbium diffusion therefore appears to be sufficient for Er clustering at 950 °C, owing to the high Er concentration in the sol-gel films (mean Er-Er distance ~ 1 nm); Er clustering generally kills the optical activity of the involved Er ions. Noting that $E_d > \Delta H$, one may estimate the probability P that Er diffusion contributes to PL deactivation rate $r_d$ from linear superposition, $E_d = P\,E_{Er} + (1–P)\,\Delta H$, which yields 0 ≤ P ≤ 0.35, where the range in P corresponds to the ± 0.7 eV uncertainty in $E_d$.

The high thermal activation energy for erbium diffusion in silica is presumably related to strong Er-O bonds, which are also responsible for stability of Er-glass structures. Although Er clustering at high temperatures is kinetically possible, a stable Si-O-Er glassy network appears energetically favoured, since the Er-O bond is strong and very stable under high temperature annealing [17] and Er acts as a primary network component at high concentrations [18]. Some notable materials that lie beyond the scope of this work are films containing the crystalline erbium compounds, such as the various Er silicates [22,37]. These are prepared with specific stoichiometries and entail a high temperature annealing step ($T$ ~ 1200 °C, also where sol-gel films are fully densified [13]) to reduce trapping by crystal defects [37].

The data in figure 3 show that strongest PL emission is obtained for annealing temperatures within a process window about 50 °C wide near $T_{max} \approx$ 850 °C, which our analysis attributes to diminished Er-OH quenching by removing OH. This observation also indicates that Er-Er CET (irrespective of whether OH is involved or not) is insufficient for quenching external emission from Er$^{+3}$ ions. One notes that the temperature range in which the Er emission begins to significantly diminish, $T >$ 900 °C, also corresponds to the region of stress relaxation in silica films [25]. Sol-gel films are usually not completely densified for anneals near 850 °C, when compared to materials subjected to high temperature annealing; we therefore attribute the apparently minimized CET near $T_{max}$ to a disordered glassy state. As OH content is progressively eliminated with higher annealing temperatures, optically active Er ions get closer together, making Er-Er CET and hence CUP processes that dissipate the excitation energy to heat become increasingly favourable. This mechanism, when coupled with possible Er-Er clustering, inevitably leads to decreased PL intensity for $T \geq 950°C$, as observed

These results are quite different from the concentration quenching observed at low Er concentrations. Existence of a range in annealing temperatures where concentration quenching may appear suppressed is a particular advantage obtained by using a high Er concentration. One does not expect thermal annealing behaviour to materially depend on Er concentration, since the optimal anneal temperature, as estimated by $T_{max}$ in equation (6), depends logarithmically on mainly the OH concentration that dominates the process. The optimum $T_{max}$ at 6 at. % Er is somewhat lower than previously reported for sol-gel films with under 1 at. % Er [3,4,15]; this is similar to the behaviour of Er-doped silicon-rich oxides, where $T_{max}$ is lowered by about 100°C at Er concentrations of 3 - 6 at. % [27].

It is instructive to compare the present results with annealing dependence of $Er^{+3}$ emission in other silica-based materials. As an example we compare (inset to figure 3) PL data obtained by Slooff et al. [9] on Er-implanted silica colloids (data from figure 3a in [9] and normalized by a factor 3.5) with the model function fitted to the sol-gel film data. Colloidal silica coatings are quite different from our spin-coated films in major respects; the low R value of the colloid process yields lower levels of bound OH, colloidal particles have larger free surface areas than sol-gel films, and there is the Er-implant damage to be annealed. Although there's much noise and uncertainty in the colloidal data, one observes a pattern of annealing behaviour that is similar to our sol-gel films. The data points suggest a shift of $T_{max}$ to lower temperature and perhaps recovery of optical activity at 1000 °C; these features may be understood in terms of our model as reflecting the lower OH levels to be removed, the greater free surface area available for OH desorption, and the lower Er concentration (0.2 at. %). Other studies of annealing behaviour have been concerned with activation of $Er^{+3}$ in deposited suboxide films ($SiO_X$:Er). For co-evaporated films Adeola et al. [27] found optimum annealing at $T_{max} \sim 600 – 700$ °C; for laser-ablation deposited films Ha et al. [28] found $T_{max} \sim 500$ °C. In these $SiO_X$:Er processes inducing $Er^{+3}$ emission entails annealing out defects and forming Si nanocrystals adjacent to Er-O complexes. Observed degradation of optical activity for annealing at high temperature ($\geq 900$ °C) is attributed in these systems to segregation effects, owing to the excess supply of Si relative to O. Significantly, $T_{max}$ for $SiO_X$:Er films are found to be lower than for Er-doped sol-gel silica films. We attribute this to the fact that OH contamination in deposited $SiO_X$:Er ought to be practically negligible.

The present sol-gel process can be suitable for creating planar amplifier devices that can be integrated with other waveguide devices, passive or active, such as splitters, switches, multiplexers or other applications related to silica-on-silicon on a single chip [38]. The same is true about the suitability of the process parameters, particularly optimum annealing temperature, for integration in silicon microelectronics.

## 6. Conclusions

A process for producing optically active $SiO_2$:Er thin films on Si substrates using low cost sol-gel techniques that utilize $Er_2O_3$ has been presented and analyzed. Photoluminescence is enhanced strongly as a function of annealing temperature, reaching optimum for annealing temperatures in a 50-°C range near $T_{max} \approx 850$ °C. External emission from Er-O structures is shielded by OH for annealing temperatures $T < T_{max}$ and by the combination of Er concentration quenching and Er-Er-OH energy diffusion for $T > T_{max}$. The results indicate that minima in quenching by Er-Er cooperative energy transfer and by Er-OH interactions can be associated with annealing near $T_{max}$.

A thermally activated reaction rate model is used to interpret the dependence of $Er^{+3}$ photoluminescence on annealing temperature. Results of this analysis yields a thermal-activation energy $E_a = 2.3 \pm 0.6$ eV for $Er^{+3}$ optical activation that is attributed to OH removal by desorption. Based on estimates of Er diffusion at high temperature, the thermal-activation energy $E_d = 2.6 \pm 0.7$ eV for $Er^{+3}$ optical de-activation is attributed to concentration quenching, residual OH and Er clustering. Thus, the model presented accurately accounts for the improved PL output and the existence of the range of optimum annealing temperatures.

## Acknowledgments

This work was supported by Deanship of Academic Research at the University of Jordan, Project contract no. 1030 and Hamdi Mango Center for Scientific Research (HMCSR) and the New Jersey Institute of Technology. One of us (SA) especially thanks Prof. N. M. Ravindra for continuous encouragement and support. Publication on this work has appeared [39].